\documentclass[a4paper,twoside]{article}
\newcommand{\affil}[1]{$^{\rm #1}$}
\usepackage[authoryear]{natbib}
\bibpunct{(}{)}{;}{a}{}{,}
\usepackage{graphicx}
\usepackage{amsmath, amssymb}
\date{} %Please leave the date blank
\title{\large\bf\flushleft The $^{13}$C Pocket In Low Mass AGB Stars.}
\author{\parbox{\textwidth}{\flushleft
\vspace{-0.5cm}
{\it O. Straniero\affil{A}\affil{C}, S. Cristallo\affil{A}, and R. Gallino\affil{B}}\\
\vspace{0.4cm}
{\small \affil{A}\,INAF-Osservatorio Astronomico di Teramo}\\
{\small \affil{B}\,Dipartimento di Fisica Generale, Universit\'a di Torino}\\
{\small \affil{C}\,Email: straniero@oa-teramo.inaf.it}}}
\begin{document}
\twocolumn[
\begin{changemargin}{.8cm}{.5cm}
\begin{minipage}{.9\textwidth}
\vspace{-1cm}
\maketitle
%
%
%%%%%%%%%%%%%     ABSTRACT    %%%%%%%%%%%%%
%Abstract of no more than 200 words here.
\small{\bf Abstract:}
It is well known that thermally pulsing Asymptotic Giant Branch stars with low mass 
play a relevant
role in the chemical evolution. They have synthesized about 30\% of the galactic carbon
and provide an important contribution to the nucleosynthesis of heavy elements (A$>$80).
The relevant nucleosynthesis site is the He-rich intermediate zone (less than $10^{-2}$ M$_\odot$),
where $\alpha(2\alpha,\gamma)^{12}$C reactions and slow neutron captures on seed nuclei 
(essentially iron) take place. A key ingredient is the interplay between nuclear processes and 
convective mixing. It is the partial overlap of internal and external convective zones that allows the
dredge-up of the material enriched in C and heavy elements. 
We review the progresses made in the last 50 years in the
comprehension of the s process in AGB stars, with special attention 
to the identification of the main neutron sources and to the particular physical conditions
allowing this important nucleosynthesis.  

%%%%%%%%%%%%%     KEYWORDS    %%%%%%%%%%%%%
\medskip{\bf Keywords:} nuclear reactions, nucleosynthesis, abundances --- stars: AGB and post-AGB 
% Please write all keywords in lower case. PASA uses the
% standard list of subject headings adopted by The Astrophysical Journal
% and available from http://www.journals.uchicago.edu/ApJ/keywords_text.html.
% Keywords are separated by em-dashes, i.e. ---

%%%%%%%%DO NOT EDIT%%%%%%%%%%%%
\medskip
\medskip
\end{minipage}
\end{changemargin}
]
\small
%%%%%%%%EDIT FROM HERE%%%%%%%%%%%%

\section{Introduction}
For a long time our understanding of the s-process nucleosynthesis and stellar evolution advanced 
in parallel, but only recently they merged in a unique theory. 
The two stories started more or less at the same time, namely in the second half
of the '50, with the first attempts to identify the main nuclear processes responsible
for the synthesis  of the elements beyond iron and with the first comprehensive 
picture of the physical properties of post-main-sequence evolutionary phases.
It was during the '70 that thermally pulsing Asymptotic Giant Branch (AGB) stars were identified 
as a promising astrophysical site for the s process, but only in the last decade of the
XX century it was realized how the so called main and strong components of the cosmic s-process
(including elements between Sr and Bi) can be produced in the He-rich intershell 
of low mass AGB stars. In this paper,
we critically analyze the main steps of these two stories up to the most recent attempts to reach a
unified theoretical scenario.   

\section{The beginning: 1950-1960}
In 1951, Bidelman \& Keenan firstly proposed a "parallel" spectral sequence of giant stars, 
including S, R, N types as well as a new possible  Ba II class. 
Stars belonging to these spectral classes show anomalously large abundances of certain 
heavy elements.  One year later, Merrill (1952) announced the discovery of technetium  
in S stars. It was a very important discovery, because Tc has no stable isotopes 
and, for this reason, 
its detection suggested that the synthesis of this element is an ongoing process 
in these stars.  

In these same years, Sandage \& Schwarzschild (1952) carried on the first models of 
Red Giant Stars and Salpeter (1952) proved that 
the $\alpha+\alpha\rightleftarrows^8$Be followed by $^8$Be$+\alpha\rightarrow^{12}$C
processes provide the way for He burning in giant stars.
The theory of stellar evolution beyond the main sequence was born.

The Cameron (1955) paper entitled "Origin of anomalous abundances of the elements in Giant Stars"
contains the first suggestion that neutron captures by iron seed nuclei are a fundamental
nucleosynthesis process. 
He identified the $^{13}$C$(\alpha,n)^{16}$O as a promising neutron source in giant stars.
A naive evolutionary scenario was depicted: as a by-product  of shell H burning 
via the CNO cycle, some $^{13}$C is left in the hot He-rich core of giant stars, 
where the $^{13}$C$(\alpha,n)^{16}$O  reaction and the consequent s-process nucleosynthesis take place.
However, Fowler, Burbidge and Burbidge (1955) argued that too few neutrons would be available
in that case (namely 1.4 neutrons per seed), because of i) the relatively low amount of $^{13}$C 
in the ashes of the CNO cycle and ii) the strong neutron poisoning of $^{14}$N, whose 
neutron capture cross section is particularly large.
A further problem concerned how these heavy elements, once produced in the 
stellar core, become observable at the stellar surface.  
In order to overcome these problems, Cameron (1957) for the first time introduced 
the need for an interplay between mixing and nucleosynthesis. 
In particular, he claimed that mixing of a few protons could take place between the
H-rich envelope and the He-rich core of low mass stars during the so called core He flash.
This mixing would allow the  production of fresh $^{13}$C 
(via the $^{12}$C$(p,\gamma)^{13}$N$(\beta^-\nu)^{13}$C chain) in the core and, at the same time, the dredge-up
of heavy elements. Actually, such a mixing occurs only if the entropy
barrier set up by the H burning shell is switched off, as happens in extremely metal poor stars
(e.g. Hollowell, Iben \& Fujimoto 1990, Picardi et al. 2004).
In the same paper, Cameron also noted that since the neutron
capture by $^{14}$N mainly proceeds through the p channel, rather than through the $\gamma$ channel, 
the neutron poisoning effect of the $^{14}$N is partially reduced to a proton recycling. 
These protons, indeed, induce an additional production of $^{13}$C. Figure 2 of the Cameron paper is 
particularly enlightening. It
 shows how a substantial overproduction ($10^3$ times the initial abundance) 
 of light-s elements (ls), like Sr,
can be obtained with a relatively low number of neutrons per seed (about 10), 
while the same overproduction of heavy s elements (hs), like Ba, and Pb require about 15 and 20
neutrons per seed, respectively.
As a matter of fact, only recently have we understood that the strong component 
of the s process most likely comes from low metallicity AGB stars, 
where the lower amount of iron favors the production of lead (e.g. Busso, Gallino \& Wasserburg, 1999).  

Another important indication can be found in Cameron's seminal paper (1957b). 
When discussing the spectra of R Andromedae, an S star,  
he explained  how the detection of  strong niobium lines, 
unlike the detection of Tc lines,
is evidence that neutron capture nucleosynthesis is not an ongoing process in this star: 
indeed, $^{93}$Nb, the only stable isotope of niobium, is destroyed by neutron capture during the AGB,
but, later on, it is produced by the delayed $^{93}$Zr decay (half-life $1.6\times10^5$ yr).
 Hence, an appreciable fraction of this
half-life has passed since neutron production ceased in the material
that has been mixed to the surface of R Andromedae.  
Later on, Cameron (1960), proposed $^{22}$Ne$(\alpha,n)^{25}$Mg as a possible alternative
neutron source. 

\section{The golden age of the classical analysis: 1961-1980}
A fundamental step ahead was taken by Clayton (1961) (see also Seeger, Fowler and Clayton 1965),
who demonstrated that a superposition of several neutron exposures is required to reproduce the 
experimentally observed abundance distribution of the s-process isotopes  
in the solar system.
As an example, Seeger et al. proposed  an exponential distribution of neutron exposures. This 
simple mathematical law was widely accepted for more than 20 years as the paradigm 
of the so called s-process classical analysis.

Meanwhile, the availability of new computer facilities in the USA, Japan  and Europe 
allowed the development of accurate models of the advanced phase of stellar evolution,
in particular the AGB phase. 
So, Weigert (1966) and Schwarzschild \& H\"arm (1967) discovered  the occurrence of thermal pulses
in the AGB phase. 
Powered by the periodic ignition of He shell flashes, a series of convective zones 
mix the products of the $3\alpha$ reactions (essentially carbon) through the He-rich layer.
Ulrich (1973) noted that,
if a potential neutron source is activated during a thermal pulse,
the partial overlap of successive convective shells naturally leads to an 
exponential distribution of neutron exposures, precisely as suggested by Clayton. 

A simple mathematical law appeared appealing for the classical analysis and, in addition,
AGB models provided a natural support to this choice. 
Then, by applying Occam's razor, the paradigm of the s process was definitely  
established. The astrophysical site of the s process was identified in the convective 
He-rich zone of  thermally pulsing  AGB stars. The resulting thermal energy is, 
in this case, about 25 KeV, corresponding to the typical maximum temperature
at the bottom of the convective zone generated by a thermal pulse (about $3\times10^8$ K).
Hence, nuclear physics investigations were concentrated on this energy range (see Bao and K\"appeler 1987). 

\section{The search of the main neutron source, successes and failures:  1980-1993}
The last piece of the puzzle, the neutron source(s), was still lacking. 
Meanwhile, the classical analysis 
grew and significant progress was made in modeling AGB stars 
(Iben 1975, Sugimoto \& 
Nomoto 1975, Truran \& Iben 1977, Iben 1981). In particular, it was realized that when the 
thermal pulse starts, $^{22}$Ne is synthesized from the $^{14}$N left behind by the H-
burning during the interpulse period, after two $\alpha$ captures and one $\beta$ 
decay. At its maximum size, the convective zone in the He-rich intershell 
extends from the position of highest He-burning rate to just below the external 
border of the H-exhausted core, but no further. Then, enough neutrons could be 
released by the $^{22}$Ne$(\alpha,n)^{25}$Mg, but only if $T\geq 3.5\times10^8$ K. 
This condition is achieved in  massive AGB only ($M>5$ M$_\odot$). Following the 
disappearance of the convective shell, the base of the convective envelope 
extends down to the outer portion of the region previously contained into the 
convective shell and allows the mixing of the material enriched in C 
and heavy elements within the envelope (the so called {\it Third dredge-up} or TDU).
   
However, the fact that the $^{22}$Ne$(\alpha,n)^{25}$Mg is very marginally active in low mass AGB stars
of nearly solar metallicity, because the maximum temperature in the He-rich zone is too low,
is at odds with the evidence that the bulk of the s-enriched AGB stars
in the disk of the Milky Way are low mass stars (Iben 1981). The prediction of extant stellar models is
that less than 1 neutron per $^{56}$Fe is produced by the $^{22}$Ne neutron source! 
The final word against the $^{22}$Ne neutron source was pronounced  by Busso et al. (1988).
On the base of stellar models of massive AGB stars, as computed by means of the FRANEC code 
in the version described by Chieffi and Straniero (1989) they concluded that the reaction  
$^{22}$Ne$(\alpha,n)^{25}$Mg is unable to give rise to a solar distribution of s-nuclei.

An alternative neutron source capable of providing a suitable neutron flux even 
in low mass AGB stars, was actually searched for since the beginning of the 1980s.
Sackmann (1980) suggested the possible occurrence of 
proton ingestion into the He convective shell
at its maximum size, but Iben (1982) showed  that no mixing of protons occurs during the 
thermal pulse because at that time the H-burning shell is very efficient, 
producing a substantial entropy barrier. On the other hand, Iben \& Renzini (1982) noted that,
as a consequence of the huge expansion of the envelope caused by the He-shell flash,
a few hundred years after the disappearance of the He-rich convective zone
the H-burning shell dies down, a condition more favorable for the downward proton diffusion.
They proposed that, at that epoch, a semiconvective layer may form, driven by an (assumed)
peak of the carbon opacity around $T=10^6$ K and that the resulting (partial) mixing may  
induce a certain proton leak from the envelope down to the top of the He-rich intershell. 
Later on, when the temperature at the top of the He-rich intershell rises, a $^{13}$C pocket may form. 
In the Iben and Renzini original scenario the s process occurred when the $^{13}$C  pocket
is engulfed into the convective zone generated by the subsequent thermal pulse 
and  it is burned at relatively high temperature through $^{13}$C$(\alpha,n)^{16}$O (see also Hollowell
and Iben 1989).
In this case, up to 26 neutrons per $^{56}$Fe are available at solar metallicity and the neutron
density is about $10^9$-$10^{10}$ neutrons per cm$^{3}$.
The first complete s-process nucleosynthesis calculation based on stellar inputs
was obtained by Gallino et al. (1988). By adopting the  Hollowell and Iben model, they
showed that the $^{13}$C$(\alpha,n)^{16}$O reaction provides, at variance with the $^{22}$Ne$(\alpha,n)^{25}$Mg,
a neutron source capable of reproducing the solar system distribution of s-only isotopes with $A>80$.
 
In spite of these successes, the crisis was, once again, knocking at the door. 
The advent of high-resolution spectroscopy era allowed powerful tools to probe the
physical conditions at the s-process site
(Smith \& Lambert 1986, Malaney 1987,  Malaney \& Lambert 1988, 
Busso et al. 1995, Lambert et al. 1995, Abia et al. 2001).
These studies used important branching ratios sensitive to the physical condition at 
the s-process site, like the temperature or the neutron density 
(see the K\"appeler et al. contribution to this volume).
In particular, the Rb/Sr ratio, which is influenced by the branching at $^{85}$Kr, is sensitive to 
the neutron density. These studies showed that the low Rb abundance, compared to Sr,
as found in many S, C and Ba stars, implies a quite low neutron density (less than $10^8$ neutrons per 
cm$^3$). This is incompatible with any s-process scenario in which the neutron irradiation 
takes place in the convective zone generated by the thermal pulse.
Similarly, the low value of the solar system isotopic ratio $^{96}$Zr/$^{90}$Zr
indicates that the main s-process nucleosynthesis  should occur at low neutron density.

Also the stellar theory was reaching a dead-end. 
Bazan \& Lattanzio (1993), investigating the energetic feedback occurring when the $^{13}$C pocket 
is engulfed into the convective shell and the s-process takes place, found likely 
substantial alterations
of the thermal pulse characteristics and those of the related nucleosynthesis. 
In particular, they found a high neutron density ($10^{10}$ cm$^{-3}$), 
implying an overproduction of both Rb and $^{96}$Zr, in clear contrast with the observation.

\section{Toward a unified approach to stellar evolution and nucleosynthesis in AGB stars}
The solution of the stellar s-process enigma was found, perhaps by chance, 
by Straniero et al. (1995). The aim was to check, by means of the FRANEC code,
the dramatic result obtained by Bazan 
and Lattanzio. 
Following the original suggestion by Iben and Renzini (1982),
Straniero et al. (1995) assumed  that after the 14th dredge-up episode of a 3 M$_\odot$ model 
of solar metallicity,
a small number of protons ($10^{-6}$ M$_\odot$) reached the top of the He-rich zone.
They then followed the evolution of these protons
through the whole interpulse period and up to the development of the 
subsequent thermal pulse.
As expected, at  H reignition a $^{13}$C pocket rapidly formed, but during the interpulse 
large enough temperature were developed for the $^{13}$C$(\alpha,n)^{16}$O reaction to be activated 
well before the end of the interpulse. In practice, the $^{13}$C  burning 
and the s process occur in radiative conditions and before the onset of the subsequent thermal pulse.
The neutron density is only $10^6$-$10^7$ cm$^{-3}$ and the typical thermal energy is just 8 KeV.
The radiative s-process timescale is quite long (a few $10^4$ yr) compared with 
the convective s-process timescale (a few years).  
New stellar models of low mass stars ($1\leq$M/M$_\odot<$3) were computed in the next 2 years,
strengthening the new scenario of the radiative $^{13}$C burning (Straniero et al. 1997).  
Gallino et al. (1998) carried out the full nucleosynthesis calculation 
based on these new models. 
The mass and the profile of $^{13}$C within the pocket were treated as free parameters.
The Gallino ST case (corresponding to $4\times10^{-6}$ M$_\odot$ of $^{13}$C)
provided the best reproduction of the main s-process component in the solar system. 
In particular, the Rb/Sr and $^{96}$Zr/$^{90}$Zr problems as well as the energetic feedback problem found by
Bazan and Lattanzio were removed. 
It is worth to note that the new stellar scenario for the s process was,
in a certain  sense, much more complex than previously believed. In fact, the different layers
within the pocket undergo different neutron exposures, because of their different 
temperature and $^{13}$C abundance. These differences are averaged 
when the s-rich pocket is engulfed into the convective zone generated by the subsequent 
thermal pulse, but due to the partial overlap with previous convective zones, the freshly
synthesized material is also mixed with the material processed by previous s-process episodes.
In addition, the
marginal activation of the $^{22}$Ne$(\alpha,n)^{25}$Mg may furtherly modify the heavy element 
distribution. As pointed out by Arlandini et al. (1999), the classical s-process analysis 
based on the simple exponential distribution of the neutron exposures 
is clearly unable to describe such a more complex nucleosynthesis scenario.   

\begin{figure}[h]
\begin{center}
\includegraphics[scale=0.30, angle=-90]{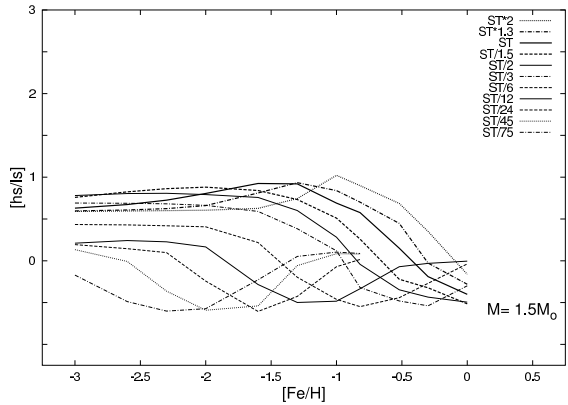}
\caption{Variation of the spectroscopic index [hs/ls] as a function of [Fe/H] for different
mass of $^{13}$C in the pocket (see text for the definition of the ST case).}\label{fig1delaude}
\end{center}
\end{figure}

More recently, Busso et al. (2001) and Abia et al. (2002) showed how the
ratios between the first (ls), the second (hs)  and the third (Pb) s-process peaks are sensitive to
the mass of $^{13}$C in the pocket. This effect is illustrated in Figure 1. 
On this basis, Gallino and his team concluded that a certain spread in the $^{13}$C mass
is required to explain the observed  spread of [hs/ls] and [Pb/hs].

\section{Conclusions}
In the last decades several hypotheses  have
been advanced to explain the origin of the $^{13}$C
pocket and quantitatively predict its mass and chemical profile,
The most promising one was that of Herwig et al. (1997, see also Herwig 2000),
where a parameterized convective overshoot was adopted,
based on prescriptions derived by 2D hydrodynamical calculations of stellar convection. They
obtained $^{13}$C pockets with mass of the order of 2-4$\times10^{-7}$ M$_\odot$
that is 10 to 20 times smaller than the ST case defined by Gallino et al. (1998).
Then, Langer et al. (1999) investigated the possibility of rotational induced mixing,
while Denissenkov \& Tout (2003) the effect of a weak turbulence induced by gravity waves. 

Let us discuss in more detail the problem of the physical description 
of the convective boundary at the time of the third dredge-up.
When the convective envelope penetrates the H-exhausted core,
a steep variation of the composition takes place at the convective boundary. The
mass fraction of H sharply drops from about 70\%, within the convective envelope, to zero, 
in the underlying radiative layer.
The discontinuity in the composition induces a sharp variation of the radiative opacity 
and, in turn, an abrupt change of the radiative gradient. 
In these conditions, the precise location of the convective border 
(i.e. the limit of the region fully mixed by convection) becomes highly uncertain.
A small perturbation causing a further mixing is amplified on a dynamical timescale, 
the radiative gradient in the radiative stable zone rises up  
and the convective instability moves toward the interior. 
This situation is commonly encountered in stellar model
computations at the time of the second and the third dredge-ups 
(Becker and Iben 1979, Castellani, Chieffi and Straniero 1990,  Frost and Lattanzio 1996,
Marconi, Castellani and Straniero 1998).
If the effect of such an instability may be maarginal in the case of the second dredge-up,
this is not true for the third dredge-up. Two are the main consequences: i) deeper dredge-ups
and ii) earlier occurrence of the first TDU episode. In other word, the minimim core mass
for the occurrence of the TDU is lower.
The observable consequences are 
many and important. The C star luminosity function is shifted toward lower luminosities and
the surface overabundance of heavy elements with A$>$80 are significantly larger.

\begin{figure}[h]
\begin{center}
\includegraphics[scale=0.40, angle=0]{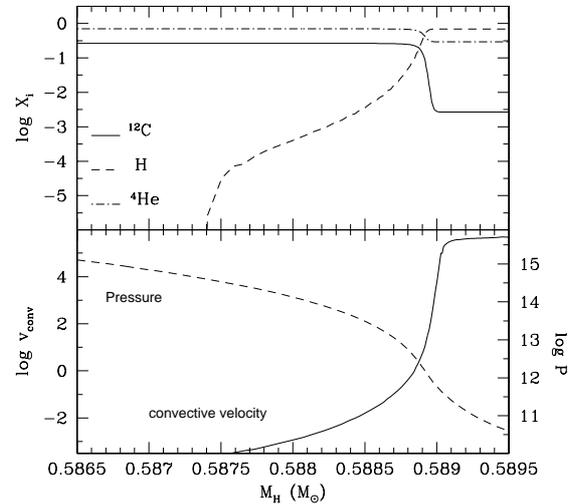}
\caption{The boundary of the convective envelope during the third dredge-up.
Lower panel: the exponential decline of the convective velocity and the sharp pressure gradient.
Upper panel: chemical composition in the transition region between the convective envelope 
and the radiative core.}\label{velco}
\end{center}
\end{figure}

A more realistic description of the convective boundary than that usually adopted in extant 
stellar evolution code is, in principle, required.  
Instead of a well defined spherical surface, as obtained when
the bare Schwarzschild's criterion is used,
the transition between the full-radiative core (i.e. unmixed) and the full-convective (i.e. fully mixed) envelope 
most likely occur in an extended zone where only a partial mixing takes place, so that a smooth and stable 
H profile may form. The evaluation of the actual extension of this transition zone and the 
degree of mixing there occurring  would require sophisticated and reliable hydrodynamical tools. 
At present, however, the inclusion of such kind of algorithms
in the hydrostatic stellar evolution codes is probably a {\it mission impossible}. 
So far, only limited, even if enlightening, hydrodynamical investigations have been carried out
(see e.g. Herwig et al. 2006). Nonetheless, a different approach may be followed.
When the convective envelope penetrates the H-exhausted core, 
the average convective velocity at the inner border 
($v$, which is proportional to the difference $\triangledown_{rad}-\triangledown_{ad}$)
rises up (see Figure 2). Note that this occurrence confirms  
the instability arising at the convective boundary during dredge-up.
In order to control such an instability, in our most recent calculations of low mass
AGB stars (Straniero et al. 2006, Cristallo et al. 2009),
we have assumed that at the convective border the average convective velocity 
drops smoothly to 0. In particular, we impose an exponential decline of $v$, 
whose steepness is given by a free parameter to be calibrated in some way:

\begin{equation} \label{param}
v=v_{bce}\exp{
\left(
-\frac{d}{\beta H_P}
\right)
} \;     ,
\end{equation}

where {\it d} is the distance from the formal convective boundary 
(as defined by the Schwarzschild's criterion),
$v_{bce}$ is the average velocity at the most internal convective mesh,
$H_P$ is the pressure scale height and $\beta$ is a free parameter. 
A similar approach has been already followed by Herwig (1997), who assumed an exponential
decline of the diffusion coefficient at the convective border.
 
The effects of the use of this algorithm on the physical and chemical properties of
the dredge-up are illustrated in Figure 2. Note that in most cases 
since $v_{bce}$ is nearly zero at a convective border, the resulting transition zone is extremely thin. 
On the contrary, during a dredge-up episode, owing to the growth of $v_{bce}$, the  
transition region, where $v$ declines smoothly to zero, becomes quite extended. 

\begin{figure}[h]
\begin{center}
\includegraphics[scale=0.35, angle=0]{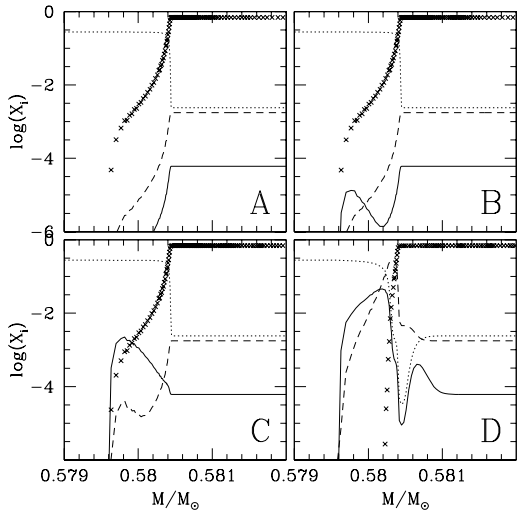}
\caption{The formation of the $^{13}$C pocket. The sequence of panels A-B-C-D shows 
the evolution of the chemical composition 
in the transition zone between the H-rich envelope and H-exhausted core.
The various lines represent: 
H (crosses), $^{12}$C (dotted), $^{13}$C (solid) and $^{14}$N (dashed).
Panel A: the third dredge up is just occurred and the convective envelope is receeding;
panel B: the temperature and the density increase, the synthesis of $^{13}$C starts; panel C:
where the number of protons is larger, some $^{14}$N is also produced; panel D:
the $^{13}$C and the $^{14}$N pockets are fully developed.}\label{tasca}
\end{center}
\end{figure}

As a result, a stable H profile is left in the He-rich intershell region after 
the third dredge-up, whose sharpness may be changed by acting on the $v$ decline parameter $\beta$.
It is in this zone that, during the interpulse, the $^{13}$C pocket forms.
The total mass of $^{13}$C depends on the $\beta$ value.
On the basis of numerical experiments made by changing the $v$ decline parameter 
we found i) the total mass of material dredged up increases with $\beta$ and ii)
there exists a maximum mass of $^{13}$C in the pocket, as obtained for $\beta$ about 0.1,
 roughly corresponding to the ST case found by Gallino et al 1998 (Figure 3). For           
 larger value of $\beta$ too much protons are mixed into the transition zone, so that $^{14}$N, rather than
 $^{13}$C, forms.        

\begin{figure}[h]
\begin{center}
\includegraphics[scale=0.40, angle=0]{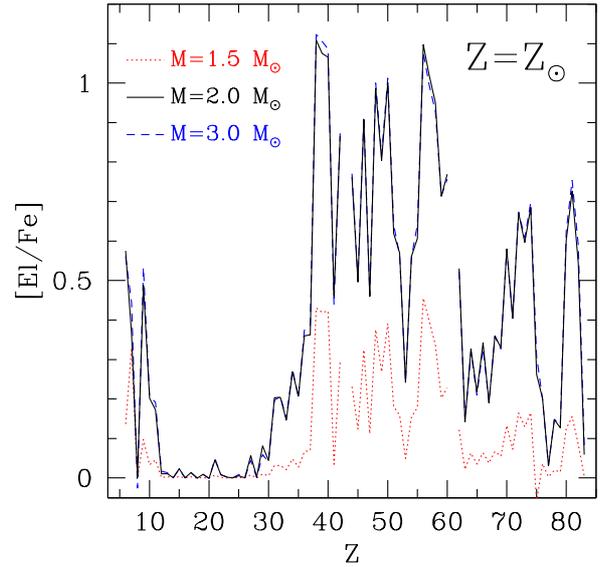}
\caption{The surface composition at the AGB tip of 3 stellar models with solar metallicity and 
$M=1.5$, 2.0 and 3.0 M$_\odot$, blue, black and red line, respectively, as obtained
by means of our latest version of the stellar evolution code, which includes
a full network fro H to Bi.}\label{chim}
\end{center}
\end{figure}
 
An important remark concerns the mixing algorithm. Our tests show that
when a diffusive scheme is adopted, as in many stellar evolution codes, the maximum resulting
$^{13}$C mass is too small to allow a sufficient s-process nucleosynthesis in the pocket (see Herwig 2000). 
On the other hand, it has been recently argued by Meakin \& Arnett (2007) that the mixing occurring 
through the convective boundaries is not a diffusive process.
As a matter of fact, $^{13}$C pockets as large as those claimed by Gallino et al (1998)
can be obtained if the degree of mixing between two (mesh) points depends 
linearly on the inverse of their reciprocal distance, as in our time dependent mixing scheme (see 
Straniero et al. 2006).

We have calculated new models of low mass stars by coupling the FRANEC stellar
evolution code with a full network, from H to Bi.
In such a way, we have been able to self-consistently derive the evolution and the nucleosynthesis
along the whole AGB. The resulting compositions at the AGB tip for 3 models of solar metallicity
are shown in Figure 4. 
The first $^{13}$C pockets, those formed after the first 2 or 3 thermal pulse episodes,
are quite large (in mass).
Later on, following the general shrinkage of the 
He-rich intershell, also the mass of the $^{13}$C pocket decreases. So, the resulting heavy
element distribution at the surface is mainly determined by the first few $^{13}$C pockets. 
Note that, owing to the overlap of the various He convective zones, 
the imprint of these first bigger $^{13}$C pockets is conserved 
up to the last dredge-up episode.  At high metallicity and in stars with 
initial mass smaller than 3 M$_\odot$, the temperature in the He
convective zone never attains values sufficiently large for the substantial activation of a 
second neutron burst driven by the $^{22}$Ne$(\alpha,n)^{25}$Mg. Decreasing the metallicity, 
thermal pulses become stronger and generate larger temperatures, so that at $Z=0.0001$
the additional contribution of the $^{22}$Ne$(\alpha,n)^{25}$Mg to the neutron capture nucleosynthesis
is rather important (Cristallo et al. 2009).

Let us conclude by noting that the theoretical scenario built up so far, after
about fifty years of intense investigations, appears to be firmly supported by the
observation of abundances in AGB stars undergoing the third dredge-up and 
belonging to different stellar populations.

\section{Acknowledgments}
This work has been supported by the italian PRIN 2006 Project
"Final Phases of Stellar Evolution,
Nucleosynthesis in Supernovae, AGB stars, Planetary Nebulae".
We are indebted with the referee for his careful reading of the paper.

%\end{multicols}


\begin{thebibliography}{aa}
\bibitem [{1}]{1}
Abia, C., Busso, M., Gallino, R., Dom\'inguez, I.,
Straniero, O., and Isern, J., 2001, ApJ, 559, 1117

\bibitem [{2}]{2}
Arlandini, C., K\"appeler, F., Wisshak, K.,
 Gallino, R., Lugaro, M., Busso, M., and Straniero, O., 1999, ApJ, 525, 886 

\bibitem [{3}]{3}
Bazan, G., and Lattanzio, J.C., 1993, ApJ, 409, 762

\bibitem [{4}]{4}
Becker, S.A., and Iben, I.J., 1979, ApJ, 232, 831

\bibitem [{5}]{5}
Bidelman, W.P., and Keenan, P.C., 1951, ApJ, 114, 473

\bibitem [{6}]{6}
Bao, Z. Y., K\"appeler, F., 1987 Atomic Data and Nuclear Data Tables, 36, 411 

\bibitem [{7}]{7}
Busso, M., Picchio, G., Gallino, R., and Chieffi, A., 1988, ApJ, 326, 196

\bibitem [{8}]{8}
Busso, M., Lambert, D.L. Beglio, L., Gallino, R., Raiteri, C.M., and Smith, V.V., 1995, ApJ, 446, 775

\bibitem [{9}]{9}
Busso, M., Gallino, R., and Wasserburg, G.J., 1999, ARAA, 37, 239

\bibitem [{10}]{10}
Cameron, A.G.W., 1955, ApJ, 121, 144

\bibitem [{11}]{11}
Cameron, A.G.W., 1957, AJ, 62, 138

\bibitem [{12}]{12}
Cameron, A.G.W., 1957b, PASP, 69, 408

\bibitem [{13}]{13}
Cameron, A.G.W., 1960, AJ 65, 485

\bibitem [{14}]{14}
Castellani, V., Chieffi, A., and Straniero O., 1991, ApJS, 74, 463

\bibitem [{15}]{15}
Clayton, D.D., 1961, Ann. Phys., 12, 331

\bibitem [{16}]{16}
Chieffi, A., and Straniero, O. 1989, ApJS, 71, 47

\bibitem [{17}]{17}
Cristallo, S., Straniero, O., Gallino, R., Piersanti, L., 
 Dom\'inguez, I., and Lederer, M.T., 2009, ApJ in press, arXiv:0902.0243 

\bibitem [{18}]{18}
Denissenkov, P.A., and  Tout, C.A., 2003, MNRAS, 340, 722 

\bibitem [{19}]{19}
Fowler, W.A., Burbidge, G.R., and Burbidge, E.M., 1955, ApJ, 122, 271

\bibitem [{20}]{20}
Frost, C., and Lattanzio, J.C., 1996, ApJ, 473, 383

\bibitem [{21}]{21}
Gallino, R., Busso, M., Picchio, G., Raiteri, C. M., and Renzini, A., 1988, ApJ, 334, 45

\bibitem [{22}]{22}
Gallino, R., Arlandini, C., Busso, M., Lugaro, M., Travaglio, C., Straniero, O., 
 Chieffi, A., and Limongi, M., 1998, ApJ, 497, 388  

\bibitem [{23}]{23}
Herwig, F., Bloecker, T., Schoenberner, D., and El Eid, M., 1997, A\&A, 324, L81

\bibitem [{24}]{24}
Herwig, F., 2000, A\&A, 360, 952

\bibitem [{25}]{25}
Herwig, F., Freytag, B., Hueckstaedt, R.M, and Timmes, F.X., 2006, ApJ, 642, 1057  

\bibitem [{26}]{26}
Hollowell, D.  and Iben, I.J., 1989, ApJ, 340, 966

\bibitem [{27}]{27}
Hollowell, D., Iben, I.J. and Fujimoto, M.Y., 1990, ApJ, 351, 245

\bibitem [{28}]{28}
Hubner, W.F., Merts, A.L., Magee, N.H., and Argo, M.F., 1977, Los Alamos Sc. Lab. Rept. (LA-6760-M)

\bibitem [{29}]{29}
Iben, I.J., 1975, ApJ, 196, 525

\bibitem [{30}]{30}
Iben, I.J., 1981, ApJ, 246, 278

\bibitem [{31}]{31}
Iben, I.J., 1982, ApJ, 260, 821

\bibitem [{32}]{32}
Iben, I.J., and Renzini A., 1982, ApJ, 263, L23

\bibitem [{33}]{33}
Iglesias, C.A., Rogers, F.J., Wilson, B.G., 1992, ApJ, 397, 717

\bibitem [{34}]{34}
Lambert, D.L., Smith, V,V., Busso, M., Gallino, R., and Straniero, O., 1995, ApJ, 450, 302

\bibitem [{35}]{35}
Langer, N., Heger, A., Wellstein, S., and Herwig, F., 1999, A\&A, 346, L37

\bibitem [{36}]{36}
Malaney, R.A., 1987, ApJ, 321, 832 

\bibitem [{37}]{37}
Malaney, R.A., and Lambert, D.L., 1988, MNRAS, 235, 695

\bibitem [{38}]{38}
Marconi, M., Castellani, V., and Straniero, O., 1998, A\&A, 160, 340

\bibitem [{39}]{39}
Meakin, C.A., and Arnett, D., 2007, ApJ, 667, 448

\bibitem [{40}]{40}
Merrill, P.W., 1952, ApJ, 116, 21

\bibitem [{41}]{41}
Picardi, I., Chieffi, A., Limongi, M., Pisanti, O., Miele, G., Mangano, G.
  and Imbriani, G., 2004, ApJ, 609, 1035

\bibitem [{42}]{42}
Sackmann, J., ApJ, 235,554

\bibitem [{43}]{43}
Salpeter, E.E., 1952, ApJ, 115, 326

\bibitem [{44}]{44}
Sandage, A.R., and Schwarzschild, M., 1952, ApJ, 116, 463 

\bibitem [{45}]{45}
Schwarzschild, M., and H\"arm, R. 1967, ApJ, 150, 961

\bibitem [{46}]{46}
Seeger, P.H., Fowler, W.A., and Clayton, D.D., 1965, ApJS, 11, 121 

\bibitem [{47}]{47}
Smith, V.V., and Lambert, D.L., 1986, ApJ, 311, 843

\bibitem [{48}]{48}
Straniero, O., Gallino, R., Busso, M., Chieffi, A., Raiteri, C.M., Limongi, M. and
   Salaris, M., 1995, ApJ, 440, L85

\bibitem [{49}]{49}
Straniero, O.,  Chieffi, A., Limongi, M., Busso, M., Gallino, R., and Arlandini, C., 1997, ApJ, 478, 332

\bibitem [{50}]{50}
Straniero, O., Gallino, R., and Cristallo, S., 2006, Nucl. Phys. A, 777, 111

\bibitem [{51}]{51}
Sugimoto D., and Nomoto, K., 1975, PASJ, 27,197

\bibitem [{52}]{52}
Truran, J., and Iben, I.J., 1977, ApJ, 216, 797

\bibitem [{53}]{53}
Ulrich, R.K., 1973, in "Explosive Nucleosnthesis", Univ. of Texas, Austin

\bibitem [{54}]{54}
Weigert, A., 1966, Z. ph. 64, 395
\end{thebibliography}
\end{document}